\documentclass[runningheads]{llncs}
\usepackage{graphicx}
\usepackage{blindtext}
\usepackage{scrextend}
\usepackage{url}
\usepackage{color, soul}
\usepackage[normalem]{ulem}
\usepackage{multirow}
\usepackage{tabularx}
\usepackage{amssymb}
\usepackage{amsmath}
\usepackage{enumitem}
\usepackage{floatrow} 
\begin{document}
\title{The MUIR Framework: Cross-Linking MOOC Resources to Enhance Discussion Forums}
\titlerunning{MUIR: Cross-Linking MOOC Resources to Enhance Discussion Forums}
\author{Ya-Hui An\inst{1,3} \and Muthu Kumar Chandresekaran\inst{1} \and Min-Yen Kan\inst{1,2} \and Yan Fu\inst{3}}
\authorrunning{An et al.}
\institute{Web IR / NLP Group (WING), National University of Singapore, Singapore\\
	\and
	Smart Systems Institute, National University of Singapore, Singapore
	\and
	Web Sciences Center, School of Computer Science and Engineering, University of Electronic Science and Technology of China, China}
\maketitle              % typeset the header of the contribution
\begin{abstract}
New learning resources are created and minted in Massive Open
Online Courses every week -- new videos, quizzes, assessments 
and discussion threads are deployed and interacted with -- in
the era of on-demand online learning.  However, these
resources are often artificially siloed between platforms and
artificial web application models.  Facilitating the linking
between such resources facilitates learning and multimodal
understanding, bettering learners' experience. 
%Muthu CR: deletin gline; abstract is usually one para
We create a framework for MOOC Uniform Identifier for Resources
(MUIR).  MUIR enables applications to refer and link to such
resources in a cross-platform way, allowing the easy minting of
identifiers to MOOC resources, akin to \#hashtags.  We demonstrate
the feasibility of this approach to the automatic identification,
linking and resolution -- a task known as Wikification -- of
learning resources mentioned on MOOC discussion forums, from a
harvested collection of 100K+ resources.  Our Wikification system
achieves a high initial rate of 54.6\% successful resolutions on key
resource mentions found in discussion forums, demonstrating the
utility of the MUIR framework.  Our analysis on this new problem
shows that context is a key factor in determining the correct
resolution of such mentions.
	
\keywords{Digital Library \and MOOC \and Learning Resource \and 
Unique Resource Identifier \and DOI \and MUIR.}
\end{abstract}
\section{Introduction} % 1 page
% Min: this is the theme for TPDL '18
Digital libraries for open knowledge goes beyond the scholarly library
and extends into the pedagogical one~\cite{mcauley2010mooc}.  While
participation in Massive Open Online Courses (MOOCs) and online
learning has expanded
\cite{hew2014students,martin2012will,pena2007giving,seely2008open},
the methods by which learners participate in these classes has still
been confined to the limitations of the Learning Management Systems
(LMS)~\cite{dalsgaard2006social,mahnegar2012learning}.  Such LMSes
often have separated and distinct views of each form of learning
resource -- discussion forums, lecture videos, problem sets, homeworks
-- where cross-linking resources is difficult or impossible to
achieve.  Learners ``cannot see the forest for the trees'' when
concepts are siloed and easy cross-referencing is impeded.

\begin{figure}[t] 
	\begin{center}
		\centering
		\includegraphics[width=\textwidth]{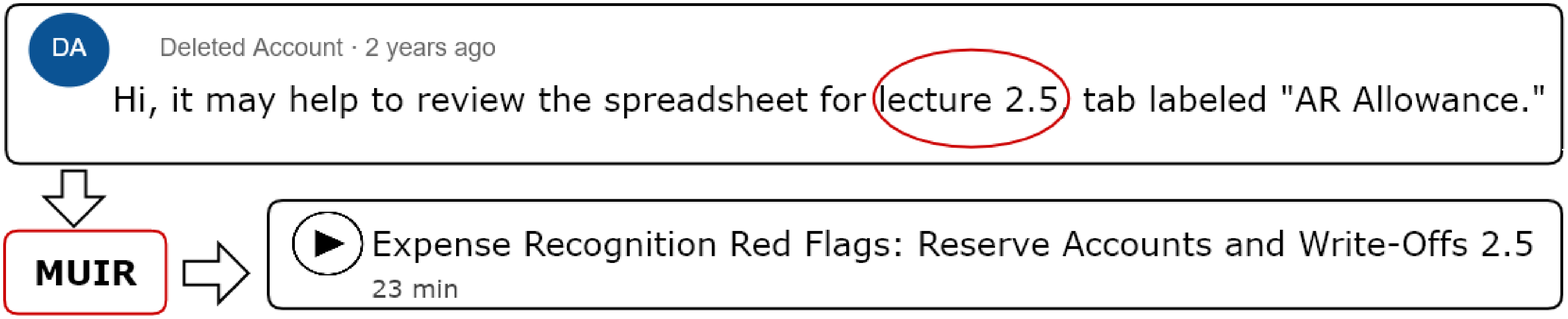}
		\caption{Crosslinking a lecture resource mention in a discussion
			forum.}
		\label{fig:introduction}
	\end{center}
\end{figure}

A concrete instance of this is in the discussion forum, where both
instructors and students co-construct arguments to support critical
thinking and knowledge~\cite{andresen2009asynchronous,marra2004content}.
Students often reference a certain quiz, this week's lecture or a
particular slide, as in Figure~\ref{fig:introduction}.  Automatically
hyperlinking such mentions to the target resource brushes and links
the two endpoints, facilitating the contextualization of course
materials across disparate views.  To address this, we introduce and
reduce to practice a pipeline that adds appropriate hyperlinks to
natural language mentions of MOOC resources in discussion forums -- a
task known as {\it Wikification}, named after the same task which was
first applied to Wikipedia.

In addressing this challenge, we needed to also propose an important
standalone contribution: a framework for MOOC Uniform Identifier for
Resources, which we name MUIR\footnote{MUIR refers to {\bf M}OOC 
{\bf U}niform {\bf I}dentifier for {\bf R}esources as well to the
eponymous framework that creates such identifiers.}.  The MUIR
framework is a two-component framework that pairs a transparent,
guessable URL syntax for learning resources with a best-effort
resolver that connects MUIR identifiers to their target resource.
Best thought of as a hybrid between bibliographic records that
identify a scholarly work, and the Digital Object Identifier that
gives a resolution, our MUIR framework facilitates the cross-linking
functionality that allows for the Wikification of natural language
mentions in learner and instructor discourse.

% MinCR: LOD can be mentioned here too. 
MUIR also facilitates resource discovery.  As a central harvester, the
MUIR resolver components crawls MOOC platforms for resources and can
expose related course material across different providers,
% YahuiCR: mention LOD here.
% MuthuCR: changed Linking Open Data to Linked Open Data
formulating a MOOC domain Linked Open Data (LOD)~\cite{bizer2009linked}, which creates typed links between data from different sources. This helps to address learning resource reuse, a problem that has been exacerbated with exponential success of MOOCs~\cite{zemsky2014mooc}. Without an aggregation service like MUIR, each MOOC LMS platform is siloed: having its own resource identifier schema that is non-portable, opaque and non-interpretable.

We demonstrate the use of the MUIR framework for the application of
Wikification. In this case study, our Wikification application
recognizes mentions to publicly exposed resources, and generates short
form references to those resources which the framework resolves and
forwards links.

\begin{figure}[t]
	\includegraphics[width=\textwidth]{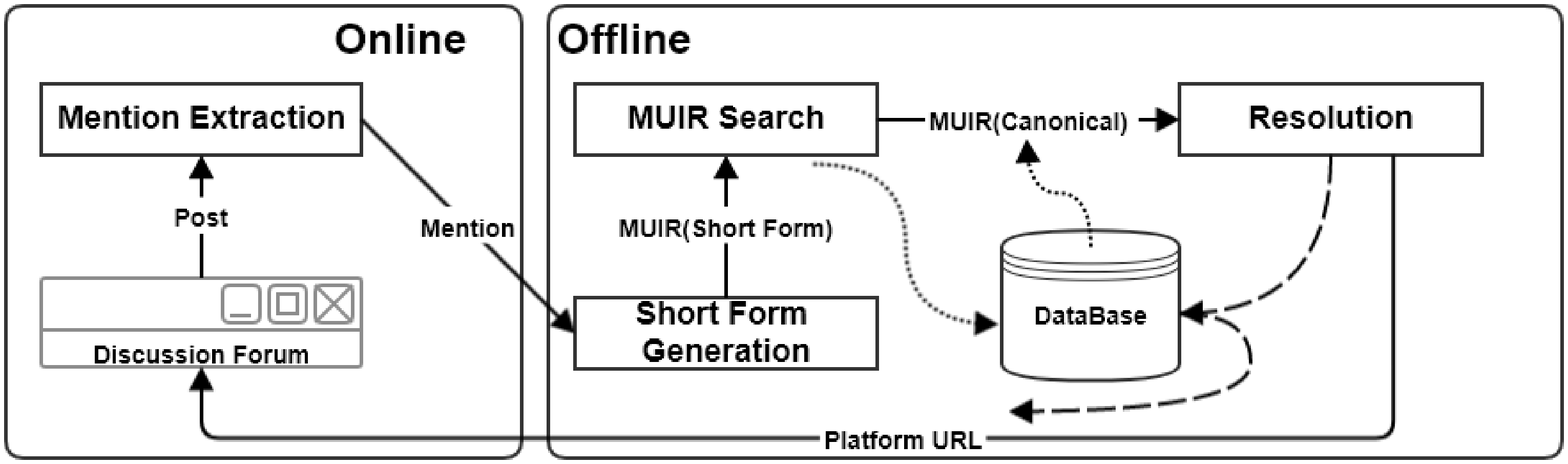}
	\caption{MUIR System Architecture: (l) online system, (r) offline harvester and resolution components.}
	\label{fig:architecture}
\end{figure}   

%%%%%%%%%%%%%%%%%%%%%%%%%%%%%%%%%%%%%%%%%%%%%%%%%%%%%%%%%%%%
\section{Related Work} % 2 pages

% MinCR: maybe you need to add a subsection or add to the second
% section on LOD.

% YahuiCR: 

The MUIR framework contributes to both the topics curation and
indexing, as well as identification schemes.  We review these areas in turn. 

\vspace{2mm}

\noindent {\bf Curation and Indexing.} Both MOOCs and Learning
Resource collection and indexing have prior work.  MOOC
List\footnote{\url{https://www.mooc-list.com/}} curates a commercial,
faceted indexing website to find current MOOC offerings.  More general
and academically inclined, MERLOT\footnote{or ``Multimedia Educational
  Resource for Learning and Online Teaching'',
  \url{https://www.merlot.org/}} achieves broader goals for thousands
of learning resources for K--12 and tertiary education, for learners,
educators, and faculty development for specific discipline.  It acts
as both an aggregator of submitted content for peer curation as well
as a focal point for gathering the community concerning these
resources~\cite{moncada2015rediscovering}. MERLOT allocates a unique
identifier to each material submitted as a pairing of a 
unique \emph{'materialId'} and an \emph{'entryType'}. More recently, the
OpenAIRE project~\cite{ameri2017exploiting} aggregates metadata about
scholarly research -- projects, publications, people, organizations,
etc.) -- into a central information space. 

\vspace{2mm}

\noindent {\bf Identifier Schemes.}  Wikification uses MUIR to
cross-link resources, creating a MOOC domain-specific form of Linked
Open Data (LOD)~\cite{bizer2009linked,mendes2011dbpedia}. It is a
method of publishing to create and publish typed links between data
entities from different sources, so that the data can be
interconnected and put to better use.  The MUIR scheme aims to
aggregate resources across platforms and should be persistent,
transparent and resolvable for various providers.  We are informed of
the the design by related resource identifiers such as PURL, DOI,
Dublin Core and general bibliographic metadata.

A Persistent Uniform Resource Locator~\cite{shafer1996introduction}
(PURL) provides a single layer of indirection built over the standard
URL protocol for web addressing.  PURLs solve the problem of
transitory URIs through their indirection, but omit any guidelines or
enforcement of the identifier minting schema; the choice of identifier
is up to the minting agent, somewhat akin to custom URL shorteners
such as {\tt bit.ly} and {\tt tiny.cc}.  The Digital Object
Identifier~\cite{paskin2010digital} (DOI) schema goes further, not
bound by any dependent protocols (e.g., HTTP for PURLs) and admits
different authorities (e.g., different journal publishers) and
distributed and hierarchical resolution via its use of the handle
system. Our MUIR proposal is technically a PURL service, where
our effort has been to create strong guidelines for the identifier
portion of the schema.

Both Dublin Core\cite{weibel1998dublin} (DC) and bibliographic
metadata are flexible containers that specify preferred (or mandatory)
metadata attribute--value fields for different types of materials,
such as {\it title} or {\it contributor}. Unlike PURL and DOI which
are opaque, MUIR opts for transparent identifiers, taking the cue from
DC and bibliographic metadata. The components of a MUIR encode the
metadata values directly as part of the URL syntax for the identifier,
and uniformly across various LMS providers.

\vspace{2mm}    

%%%%%%%%%%%%%%%%%%%%%%%%%%%%%%%%%%%%%%%%%%%%%%%%%%%%%%%%%%%%
\section{The MUIR Framework} % 2 pages
\label{framework}

\noindent ``When we try to pick out anything by itself, we find it
hitched to everything else in the universe'' --- John Muir \\

Our uniform identifier scheme for MOOC learning resources, embodies
the American naturalist John Muir's insight that everything is
interconnected.  In creating MUIR, our aim is to objectify MOOC
resources so that they can be inventoried, referenced and subsequently
better ``hitched'' to other resources, in the spirit of LOD, creating
a densely tangled web of knowledge crucial for the contextualization
of learning.  We discuss the desiderata for our MUIR schema, while
relating it to the practices of related work.

We motivate this section by working through the elements of a
hypothetical MUIR associated with a learning resource from Coursera
representing a specific lecture on accounting analytics:

\begin{figure}[t]
  \fbox{
    \begin{minipage}[t]{0.95\textwidth}
      \small
      \noindent {\bf I. MUIR (Short Form, Transparent (sample)): } \tt www.example.org/ accounting-analytics/Week 2/lecture/2-5 \vspace{2mm} \\
      \noindent {\bf II. MUIR (Canonical Transparent): } \tt www.example.org/Coursera/accounting- analytics/1480320000000/Brian J Bushee\&Christopher D. Ittner/Videos/
      expense-recognition-red-flags-reserve-accounts-and-write-offs-2-5 \vspace{2mm} \\
      \noindent {\bf III. MUIR (Opaque): } \tt www.example.org/id/1239jdn3oni3123s \vspace{2mm} \\
      \noindent {\bf IV. Coursera URL:} \tt www.coursera.org/learn/accounting-analytics/lecture/ 1UzkX/expense-recognition-red-flags-reserve-accounts-and-write-offs-2-5 
    \end{minipage}
  }
  \caption{A Coursera learning resource URI in MUIR's threefold
    identifier scheme.}
  \label{fig:runningexample}
\end{figure}

\vspace{2mm}
\noindent {\bf 1. Indirection.}  MUIRs provide two layers of
indirection over actual resolvable resources such as a Coursera
discussion forum, or a quiz hosted on a course on EdX.  The first
layer serves as a semantically transparent, short form where fields
can be omitted and the search functionality of MUIR invoked to form
the best-effort resolution to the canonical form. Similar to the
simplicity of \#hashtags, the MUIR short form encourages direct use by
humans, later to be resolved to a canonical form or directly to the
platform URL via best-guess relevance search.

The second layer of indirection (from the canonical form to the
platform URL) provides both a uniform access mechanism to
the resources that is platform- / provider-independent.  As with
PURLs, it also lends itself to preservation, having a single authority
for resolution.  Both the canonical form and the opaque form map one
to one to the platform instance. \vspace{2mm}

\noindent {\bf 2. Transparent.}  Unlike traditional schema that use
succinct opaque identifiers to serialize and identify objects, MUIR
takes the cue from bibliographic systems that admit multiple,
value--attribute fields to name resources. Much like how Dublin Core
mandates certain fields be specified, MUIR also splits fields into
required ({\it Resource Title}, {\it Resource Type}, {\it Course
	Name}, {\it Session Date}, {\it Instructor(s)}, {\it
	Institution}, {\it Source Platform}) and optional categories ({\it
	Other Elements}).  The short form MUIR invokes search by the
resolution system to find the most appropriate learning resource, akin
to search in a web search engine or an online public access catalog.
\vspace{2mm}

\noindent {\bf 3. Comprehensive.}  MUIR's resource type categorizes
the most common learning resources exposed in MOOCs.  We survey
learning resources provided on 29 worldwide MOOC platforms to
inventory the common learning resources exposed, and map these forms
to MUIR's {\it Resource Type} (Table~\ref{table:moocs}).
{\it Videos} present the lecture content.  {\it Slides} provide the
lecture content for download and separate review, often aligned to
those in the video.  {\it Transcripts} of the videos are sometimes
available for various languages, often for other languages than the
one used in the video.  {\it Assessments} capture any form of
assessments, exercises, homeworks and assignments that aim to
self-diagnose the learners' knowledge commitment of the course
content.  {\it Exams} evaluate the knowledge and/or skills of
students, including quizzes, tests, mid-exams and final examinations.
{\it Readings} optionally provide a list of other learning resources
provided by courses.  {\it Additional Resources} help to catch other
materials made available for specialized discipline-specific
courses. For example, computer programming courses can provide program
files for reference.

\begin{table}[t]
	\caption{Prevalence of resource types exposed on global MOOC platforms. `Scale' indicates \# of courses / \# of learners.  The subsequent columns on top represents videos, slides, exams, quizzes, transcript, homeworks, assignments, assessments, exercises, readings, articles, programming scripts and additional materials, respectively. Each resource type is mapped to one of MUIR's canonical resource types (bottom row).}
	\label{table:moocs}
	\scriptsize
	\begin{tabular}{|l|c|rl|l|c|l|l|l|l|c|l|l|l|l|c|l|l|}
		\hline
		
		No. Platform& Country &Scale (C&/L) &  \rotatebox{90}{V.} & \rotatebox{90}{S.} & \rotatebox{90}{E.} & \rotatebox{90}{Q.} & \rotatebox{90}{Tr.} &\rotatebox{90}{HW.} &\rotatebox{90}{Asg.} & \rotatebox{90}{Ass.} &\rotatebox{90}{Ex.} & \rotatebox{90}{Re.} & \rotatebox{90}{Art.} & \rotatebox{90}{Pro.}  & \rotatebox{90}{Add.}\\ \hline
		
		1. Coursera&US&2000+&/25M+ &$\checkmark$&$\checkmark$&$\checkmark$&$\checkmark$&$\checkmark$&$\checkmark$&$\checkmark$&$\checkmark$&$\checkmark$&&&$\checkmark$&$\checkmark$ \\ \hline
		2. edX&US&950+&/14M+
		&$\checkmark$&$\checkmark$&$\checkmark$&$\checkmark$&$\checkmark$&$\checkmark$&$\checkmark$&$\checkmark$&&$\checkmark$&&$\checkmark$ &$\checkmark$  \\ \hline
		3. Udacity&US&200+&/4M+
		&$\checkmark$&&&$\checkmark$&$\checkmark$&&&&&&&&  \\ \hline
		4. FutureLearn&UK&400+&/6.5M+
		&$\checkmark$&&&$\checkmark$&&&$\checkmark$&&$\checkmark$&&$\checkmark$&&$\checkmark$  \\ \hline
		5. iversity&GER&50+&/0.75M+
		&$\checkmark$&  &&&&&&$\checkmark$&&&&&\\ \hline
		6. Open2Study&AU&45+&/1.1M+
		&$\checkmark$&$\checkmark$&&$\checkmark$&$\checkmark$&&&$\checkmark$&&&&&  \\ \hline
		7. Acumen+&US&34+&/0.3M
		&$\checkmark$&&&&$\checkmark$&&$\checkmark$&&&$\checkmark$&&&$\checkmark$  \\ \hline
		8. P2PU&US&200+&/---
		&$\checkmark$&$\checkmark$&$\checkmark$&$\checkmark$&$\checkmark$&$\checkmark$&$\checkmark$&$\checkmark$&$\checkmark$&&&$\checkmark$&$\checkmark$  \\ \hline
		9. Academic Earth&US&600+&/5.8M+&
		$\checkmark$&$\checkmark$&$\checkmark$&$\checkmark$&$\checkmark$&$\checkmark$&$\checkmark$&$\checkmark$&$\checkmark$&$\checkmark$&$\checkmark$&$\checkmark$&$\checkmark$  \\ \hline
		10. Alison&IE&1000+&/11M+
		&$\checkmark$&&&&&&&&&&&&  \\ \hline
		11. Athlete&CH&27+&/14K+
		&$\checkmark$&&&&$\checkmark$&&&&&$\checkmark$&&& \\ 
		Learning Gateway&&&&&&&&&&&&&&&&\\\hline
		12. Canvas Network&US&200+&/0.2M+&$\checkmark$&&&$\checkmark$&$\checkmark$&&$\checkmark$&&&&&& \\ \hline
		13. Course Sites&US&493+&/---&$\checkmark$&&&$\checkmark$&&&$\checkmark$&&&$\checkmark$&&&$\checkmark$ \\ \hline
		14. KhanAcademy&US&---&/57M+&$\checkmark$&&&&$\checkmark$&$\checkmark$&&$\checkmark$&$\checkmark$&&&& \\ \hline
		15. Open Learning&JP&30+&/---&$\checkmark$&$\checkmark$&$\checkmark$&$\checkmark$&$\checkmark$&$\checkmark$&$\checkmark$&$\checkmark$&$\checkmark$&$\checkmark$&$\checkmark$&$\checkmark$&$\checkmark$ \\ \hline
		16. OpenupEd&EU&190+&/---&$\checkmark$&$\checkmark$&$\checkmark$&$\checkmark$&$\checkmark$&$\checkmark$&$\checkmark$&$\checkmark$&$\checkmark$&$\checkmark$&$\checkmark$&$\checkmark$&$\checkmark$ \\ \hline
		17. Saylor&US&100+&/---&$\checkmark$&&$\checkmark$&$\checkmark$&&&&&$\checkmark$&&$\checkmark$&$\checkmark$& \\ \hline
		18. Udemy&US&---&/20M+&$\checkmark$&&&$\checkmark$&&&&&&&&&$\checkmark$ \\ \hline
		19. CNMOOC&CN&600+&/---&$\checkmark$&&&$\checkmark$&&&&&&&&& \\ \hline
		20. Complexity &US&11+&/---&$\checkmark$&&&$\checkmark$&$\checkmark$&$\checkmark$&&&&&&&$\checkmark$\\ 
		Explorer & & & & & & & & & & & & & & & & \\ \hline
		21. Ewant&TW&600+&/20K+
		&$\checkmark$&$\checkmark$&&$\checkmark$&&&&&&&&& \\ \hline
		22. Janux&US&20+&/31K+
		&$\checkmark$&&&$\checkmark$&$\checkmark$&&$\checkmark$&&&&&&$\checkmark$ \\ \hline
		23. Microsoft&US& 800+&/---
		&$\checkmark$&$\checkmark$&&$\checkmark$&$\checkmark$&&&&&&&& \\ 
		Virtual Academy &&&&&&&&&&&&&&&&\\ \hline
		24. NTHU MOOCs&TW&46&/---
		&$\checkmark$&&&$\checkmark$&&&$\checkmark$&&&&&& \\ \hline
		25. Stanford Online&US&100+&/---
		&$\checkmark$&$\checkmark$&$\checkmark$&$\checkmark$&&$\checkmark$&$\checkmark$&$\checkmark$&&&&&$\checkmark$ \\ \hline
		26. XuetangX&CN&1300+&/9M+
		&$\checkmark$&$\checkmark$&$\checkmark$&$\checkmark$&&$\checkmark$&$\checkmark$&$\checkmark$&&&&&$\checkmark$ \\  \hline
		27. icourse163&CN&1000+&/---
		&$\checkmark$&&$\checkmark$&$\checkmark$&&$\checkmark$&$\checkmark$&&&&&& \\ \hline
		28. FUN&FR&330+&/1M+
		&$\checkmark$&&&$\checkmark$&&&&&$\checkmark$&&&& \\\hline
		29. FX Academy&ZA&10+&/---
		&$\checkmark$&&&$\checkmark$&&&&&&$\checkmark$&&&$\checkmark$ \\ \hline \hline
		\multicolumn{4}{|l|}{\bf \# of platforms w/ Resource Type} &29&11&10&24&15&11&16&11&9&8&5&6&15 \\ \hline
		\hline
		\multicolumn{4}{|l|}{{\bf Mapping to MUIR's Resource Type}}&V.&S.&\multicolumn{2}{c|}{E.}&T.&\multicolumn{4}{c}{Ass.}&\multicolumn{2}{|c}{Re.}& \multicolumn{2}{|c|}{Add.} \\ \hline
		
	\end{tabular}
\end{table}
\vspace{2mm}		
\noindent {\bf 4. Stable. } In addition to standard descriptor-like
identifier structure, MUIR also has an alternate serial identifier
syntax that is opaque and succinct, permitting short references that
are permanent, as in the final MUIR opaque identifier in
Figure~\ref{fig:runningexample}.  Thus there can be many MUIR
short form, transparent descriptors that map to a single unique opaque identifier.

\subsection{Collected Dataset}\label{section-data}

We operationalize our MUIR framework by creating a series of crawlers
to proactively collect learning resources from MOOC platforms.  In the
remainder of the paper, we study using MUIR against a subset of
crawled resources from Coursera as a proof of concept.  Our Coursera
corpus, collected at January 31, 2017, includes all posts and
resources of 142 courses that had already completed, totalling 
102,661 posts and 11,484 learning resources spanning all 7 resource 
types.

%%%%%%%%%%%%%%%%%%%%%%%%%%%%%%%%%%%%%%%%%%%%%%%%%%%%%%%%%%%%
\section{Discussion Forum Wikification}
%Muthu: global: Why is Wikificaiton initcapitalised?
% Yahui: The word 'Wikification' derives from 'Wikipedia', thus I use it initially capitalised.  
% Yahui: The Discussion Forum Wikification is a system, thus I change back the section name to 'Discussion Forum Wikification' 
% MinCR: Wikification with an initial cap should be correct
We operationalise the MUIR framework through the task of {\it discussion
	forum Wikification}. Our system for forum Wikification extracts 
and hyperlinks mentions of learning resources in student posts as
shown in Figure~\ref{fig:introduction}.

The skeptic might ask: Is Wikification meeting a real demand for
crosslinking learning resources?  To answer this, we wish to calculate
the number of mentions that are actually present in discussion forum.
Let us assume that mentions to the seven resource types do contain a
descriptive keyword.  While the presence of these keywords may not
necessarily denote an actual mention (i.e., ``{\it I have a
	question}''), the percentage of posts that contain the relevant
keywords serves as an upper bound for the number of mentions.
Restricting our examination to content subforums (excluding forums for
socializing; e.g. `{\it Meet \& Greet}' and `{\it General
	Discussion}'), we find that approximately 15,529/69,025 = 22.5\%
posts contain one or more keywords. Restating, about 1 of 4 posts in
discussion forums potentially have mentions that need Wikification.
So there is a real need that we address with Wikification.

% Camera-ready version - Yahui: 4 concrete steps
The process presumes that the MUIR system has proactively crawled and
indexed MOOC resources, as previously discussed. We reduce the problem
into 4 concrete phases as shown in Figure~\ref{fig:architecture}: 1)
Mention Extraction: mention identification, 2) Short Form Generation:
MUIR short form construction, 3) MUIR Search: MUIR short form to
canonical form resolution, 4) Resolution: forwarding the request to
the platform URL.  Note that the first two phases take place outside
of the MUIR framework, in our Wikification application that processes
discussion forums.  We step through these four phases in turn to
illustrate how the MUIR framework interacts with the Wikification
process.
\vspace{-2mm}
\vspace{5mm}

{\bf Phase 1: Mention Extraction. } Wikification begins by identifying
important mentions from a post of a course.  As natural language
mentions can occur in an infinite variety, in this initial study, we
constrain the problem scope to identifying only {\bf S}ingle, {\bf
  C}oncrete, w{\bf I}thin-course entities (or SCI).  As
counterexamples, references to collective entities (i.e., ``the
quizzes''), specific topics taught within a course (similar to
keywords, i.e., ``corporate risk'') fall outside the scope of our SCI
definition.

Analyzing actual SCI mentions in discussion forums, such as
\emph{``lecture 2.5''} in Figure~\ref{fig:introduction} and those in
Figure~\ref{fig:mentions-sample} show us that SCI entities do lend
themselves to be captured by a simple regular expression matching with
a keyword followed by a numeric offset.  We thus programmatically find
and delimit such mentions as spans for hyperlinking. This solution,
although overly simplistic, serves well as a starting point for
Wikification.  We revisit this decision later in our evaluation.
\vspace{2mm}

{\bf Phase 2: Short Form Generation.}  For each mention, Wikification
generates a MUIR short form programmatically. The short form is used
to split the mention into component words, using which our algorithm
maps them to fields in the MUIR short form.  Inferrable missing
components are added by the context of the hosted discussion
forum. Continuing with our running example, this stage takes the
mention \emph{``lecture 2.5''} that appears in an {\it Accounting
	Analytics} course on Coursera, and constructs the short form I in
Figure~\ref{fig:runningexample}, where the mention's text of
\{\emph{``lecture''}, \emph{``2''}, \emph{``.''} and \emph{``5''}\}
constructs the $s_4$ and $s_5$ short form components: the relative
block number ($2{-}5$ denotes module 2 lecture 5), and remaining
components ($s_2$ and $s_3$) are inferred from context:
\vspace{-2mm}
$$ \underbrace{\mathtt{www.example.org}}_{s_1}/\underbrace{\mathtt{accounting{-}analytics}}_{s_2}/\underbrace{\mathtt{Week 2}}_{s_3}/\underbrace{\mathtt{lecture}}_{s_4}/\underbrace{\mathtt{2{-}5}}_{s_5} $$

\noindent Here, $s_1$ is the MUIR resolver host, $s_2$ is the course
name, $s_3$ is the forum name (usually the week number) of the post,
$s_4$ is the resource type and $s_5$ represents the relative block
number.
\vspace{2mm}

{\bf Phase 3: MUIR Search.}  A click on a short form requests the
resource from MUIR resolver. This search process is the first layer of
indirection, combining the post information in the MUIR database from
which MUIR obtains additional peripheral information (platform,
session date and instructor(s) name) about the post that embeds the
mention.
The search process first utilizes the origin post data \{source
platform, $s_2$, session date and instructor(s) name\} to locate the
hosting course's context. The remainder of the short form ($s_4$ and
$s_5$) are used to match the resource type and name in a full text
search, where exact matches are favored. 
The resolver searches its index of canonical MUIRs using
this custom search logic to match with
the short form and deems the best match its resolution. As in the running
example, this process matches the MUIR short form I to the MUIR
canonical form II: 

\vspace{-5mm}
\scriptsize
$$\underbrace{\mathtt{www.example.org}}_{f_1}/\underbrace{\mathtt{Coursera}}_{f_2}/\underbrace{\mathtt{accounting- analytics}}_{f_3}/\underbrace{\mathtt{1480320000000}}_{f_4}/\underbrace{\mathtt{Brian J Bushee\&Christopher}}_{f_5}$$
\vspace{-5mm}
$$\underbrace{\mathtt{D. Ittner}}_{f_5}/\underbrace{\mathtt{Videos}}_{f_6}/\underbrace{\mathtt{expense{-}recognition{-}red\-flags{-}reserve{-}accounts{-}and{-}write{-}offs{-}2{-}5}}_{f_7} $$
\normalsize

\noindent Here, $f_1$, $f_3$ and $f_6$ are migrated from the short
form, and the remaining fields have been imputed from context: $f_2$,
$f_4$, $f_5$ and $f_7$ give the source platform, the session date,
instructors' names, and the slug name of the resource, respectively.

\vspace{2mm}		
{\bf Phase 4: Resolution.}
This final phase is simple, as the canonical MUIR maps one-to-one with
a platform URL, through a hash table lookup. This process maps the
running example's canonical form II to the platform-specific URL IV
through the second layer of indirection.

%%%%%%%%%%%%%%%%%%%%%%%%%%%%%%%%%%%%%%%%%%%%%%%%%%%%%%%%%%%%
\section{Wikification Evaluation} % 2 pages
\label{experiment-results}

We believe the MUIR identifier framework is useful on its own right,
but it is hard to evaluate its intrinsic utility.  We instead evaluate
extrinsically, assessing the utility of MUIR as a component within
discussion forum Wikification.  Specifically we ask ourselves the
following research questions (RQ):

\begin{description}
	\item[RQ1.] What is the coverage rate for posts that actually contains
	mentions?
	\item[RQ2.] How accurate is the resolution for different resource types?
\end{description}

%% We use two metrics in evaluation: {\it coverage} and {\it
%%   recall}. Recall measures the performance of searching relevant
%% resources, calculating the ratio of accurately resolved mentions to
%% the total number of mentions.

%% We use coverage to one is the ratio of posts
%% contains mentions, and the other is the ratio of mentions extracted by
%% the simple pattern we use comparing with the mentions actually exists
%% regardless of successfully extracted or not.

\noindent{\bf RQ1: Mention Coverage.}  With a full annotation of the
dataset we could conclusively measure the coverage of our regular
expressions in capturing actual natural language mentions to
SCI. However, the effort for full annotation is infeasible, and
instead we randomly sample $\sim$1,000 posts to check the actual
coverage of our Wikifier syntax.  We note that it can be unintuitive
for annotators to identify whether a word, phrase or sentence is a
mention, so we employed two independent annotators to reduce
bias. Results for this sample annotation are shown in
Table~\ref{tab-mention-coverage}.

\begin{table}
	\caption{Mention extraction coverage.}\label{tab-mention-coverage}
	\begin{center}
		\begin{tabular}{ | l | c | c | c | c |c | } \hline
			Annotator 	& \# of  & \# of posts identified 	&  \# Extracted by  & \# & \\ 
			ID & Posts & as having mentions & our Wikifier & Correct & Coverage \\ \hline 
			Annotator 1	& 1,087	& 156	& 5	& 5	& 14.4\% \\ \hline
			Annotator 2	& 1,087	& 175	& 5	& 5	& 16.1\% \\ \hline \hline
			Overall 	& 1,087	& 196 (Union)	& 5	& 5	& 18.0\% \\ \hline
		\end{tabular}
	\end{center}
\end{table} 

In our 1K sample of posts, $18\%$ of posts or more contain mentions to
learning materials.  This is significant, as it shows that there is
much potential to better interlink resources, even just for the silo
of discussion forums. In these sampled posts, our Wikifier matched 5
mentions, which were all actual mentions (correct).  This result shows
that our $<${``\it keyword}'' + number$>$ pattern has high precision
but suffers from low recall, covering only about $2.6\%$ of possible
mentions.

How can we improve mention extraction coverage?  We examine the causes
for the coverage disparity, where the parenthetical percentage is
determined over the same sampled data.

\noindent
\begin{figure}[t]
	\fbox{
		\begin{minipage}[t]{0.95\textwidth}
			\textbf{YES:} $\langle m_1 \rangle$ Is it just me or were some questions on \uline{Quiz 2} a surprise? There were a few questions that were not discussed in the lesson plan.\\ \small
			\textbf{YES:} $\langle m_2 \rangle$ Hello, I just would like to note that on 12:30 in the answer to question 3 in the \uline{lecture 2.4} it says that the network is deadlock-free, whereas ...\\ % counter example: if t1 and t2 are executed, there is only one token in p3, thus t3 cannot be activated and the network is in a deadlock. 
			\textbf{NO:} $\langle m_3 \rangle$ The last item, that is ``\uline{Probability Models for Customer-Base Analysis.pdf}'', in the Resources \&gt; Additional Readings by Week section for Week 3 is not accessible. \\ \small
			\textbf{NO:} $\langle m_4 \rangle$ I'm working on the \uline{programming assignment for ML}, week 2.  I successfully submitted answers to the obligatory questions.\\ \small
			\textbf{NO:} $\langle m_5 \rangle$ At \uline{around 5:00 in the lecture}, we see that the regularization term in the cost function is summed from 1 to L-1.  Shouldn't this be 2 to L?\\ \small
			\textbf{NO:} $\langle m_6 \rangle$ Hello. I wanted to use ``e'' as a number for \uline{ex.2/week3}. It didn't work, and I didn't find useful help with ``help exponent''.
		\end{minipage}
	}
	\caption{Actual resource mentions in our 1,087 sample sized dataset,
		illustrating the variety of expressions.  Our Wikification
		currently handles the first two mentions.}
	\label{fig:mentions-sample}
\end{figure}

\begin{enumerate}
	\item{\bf Implicit Contextual Knowledge ($\sim$45\% of errors).} In
	sequential posts, posters often refer to the content from the
	previous posters, and refer using demonstrative pronouns such as
	`{\it this}', `{\it that}' or `{\it the}'.  Without context
	knowledge, our prototype simply does not capture such mentions, such
	as in `{\it that video you mentioned}'.
	\item{\bf Named Reference ($\sim$30\% of errors).} Direct use of the
	resource name -- especially for videos, slides and quizzes -- makes
	such mentions impossible to capture, without predicating prior MUIR
	lookup (cf $m_3$ in Figure~\ref{fig:mentions-sample} or `{\it the
		problem ``Hashing with chains''}').
	\item{\bf Informal Expressions ($\sim$15\% of errors).} Colloquial
	expressions abound (Figure \ref{fig:mentions-sample}'s $m_4$ and
	$m_6$) and fall outside the current scheme.  Adding regular
	expressions to capture these would improve coverage at the cost of
	precision.
\end{enumerate}

{\bf RQ2. MUIR Resolution Accuracy.}  The other component that needs
evaluation is Phase 3, MUIR Search.  Given the short forms that are
generated by Wikification, MUIR Search connects the short form to a
(hopefully correct) platform URL.

\begin{table}[t]
	\caption{Resolution Accuracy Evaluation.  Only mentions to 4 MUIR
		types are present in our Coursera subset.  P\_I represents precision of Annotation I and P\_II is for Annotation II.}
	\label{tab:annot-1}
	\begin{tabular}{|l|c|c|c|}\hline
		Resource & \# of instances & P\_I  & P\_II \\  \hline
		Videos  & 89	& 71.9\% & 57.3\% \\ \hline
		Slides  & 27	& 74.1\% & 33.3\% \\ \hline
		Exams   & 718  & 83.0\%	& 53.3\% \\ \hline
		Assessments & 12 & 50.0\% & 25.0\% \\ \hline
		Total   & 846  & 81.1\% & 54.6\% \\ \hline 
	\end{tabular}
\end{table}

% We offer two evaluations that give evidence on the resolution precisions of our custom search logic, shown in Table ~\ref{tab:annot-1}. In the first evaluation (P\_I), we examine the search logic on data that generated by only depending on the related data of mentions, just as machine does. This gives a pessimistic upper-bound sense for how well the proposed search logic can do. In the second (P\_II for Annotation\_II), we also exhaustively annotated the ground truth test data by considering all of the context of the mentions including the content around the mentions and other posts in the same thread to properly resolve. We then check whether MUIR search resolve in reality.

We offer two evaluations that give complementary data on the
resolution accuracy, shown in Table~\ref{tab:annot-1}. Comparing P\_I
against P\_II, the accuracy of Annotation~I is generally better than
Annotation~II. That is because Annotation~I % (the first test dataset)
is generated only by depending on the information of mentions and the
limited relevant information of posts, foregoing the implicit
contextual knowledge of the previous and subsequent posts.  This gives
an upper-bound for how well mentions are actually resolved by our
simple search logic. But in Annotation~II %(the second test dataset),
when we annotate the ground truth test data, we consider all of the
context of the mentions including the content around the mentions and
other posts in the same thread. This is a realistic evaluation on the
full complexity of the problem.

The results are best analyzed jointly.  We see that the mentions we
capture are easy to extract (higher performance on Annotation~I), but
hard to resolve without context (lower performance on
Annotation~II). The accuracies for four {\it Resource Types} have
different degrees of reduction. But the results are encouraging: our
prototype, even with its simple logic, can already handle almost 55\%
of learning resources.

As we did for RQ1, we further categorized a rough cause to the errors
in the resolution process:
% \vspace{-1mm}
\begin{enumerate}
	\item{\bf Mentions needing context to resolve against multiple matches
		($\sim20\%$ of errors):} Learners may write mentions such as ``{\it
		lecture 4.5}'', where ``{\it 4}'' and ``{\it 5}'' are used by MUIR
	Search but could refer to different lectures that both have textual
	components ``4'' and ``5'' in their slug name.
	\item{\bf Multiple potential targets ($\sim70\%$ of errors):} Even
	considering context, certain mentions are still ambiguous.  If a
	mention states ``{\it question 3}'' but there are multiple quizzes
	within the context, all which have a Question 3, the target is
	ambiguous.  MUIR can only guess in this case.
	\item{\bf Errors in mention extraction ($\sim10\%$ of errors):} These
	are cascaded from the Phase 1 process of mention
	extraction. Examples include {\it partial mention extraction}
	(``{\it lecture's 2 transcript}'' may be written by a learner, but
	only ``lecture 2'' was detected) and {\it informal reference} ({\it
		cf} $m_6$ in Figure~\ref{fig:mentions-sample}).
	%%Muthu: Proofread until here
	% \item{\bf Ambiguous keyword ($\sim00\%$ of errors):} With mentions
	%   such as ``{\it module 3}'', the word ``{\it module}'' could
	%   variously refer to the encapsulating class Week 3 resource or to a
	%   resource named ``{\it module 3}''.
\end{enumerate}

%For the current system, the reason why the accuracy for ground truth
%test data is lower than Annotation I are mainly reflected in the
%following aspects:

Both RQ1 and RQ2 discussions clearly point forward in the direction of
improving coverage, especially in Phase 1, as such errors cascade.  A
clear direction is to incorporate contextual knowledge: our current
work thus aims to incorporate such knowledge by the machine reading of
the posts, by leveraging recurrent neural network based learning
models~\cite{sutskever2014sequence} currently making much impact in
natural language processing research.  This will help the Wikification
process by both capturing more natural mention expressions and minting
better Phase II MUIR short forms that better facilitate correct
resolution downstream.

We note that mention extraction can also be facilitated by introducing
linking conventions, similar to \#hashtags.  MUIR's short form can be
further facilitated by the future learner's explicit triggering when
writing their posts: i.e., ``{\it I have a question about
	\underline{\#video5}}'', where mention identification are solved by
the learner.
%%%%%%%%%%%%%%%%%%%%%%%%%%%%%%%%%%%%%%%%%%%%%%%%%%%%%%%%%%%%
%\section{Conclusion} % .5 pages
\section{Conclusion}
\label{conclusion}
For a learner to see the forest for the trees requires seamless
interlinking of learning resources.  Discussion forum Wikification
takes us closer towards this goal. Our prototype shows
the feasibility of the approach for simple mention types, and further
motivates research on better mention identification and search
resolution of such mentions.

Underlying this development is our core contribution of the MUIR
framework for identifying and referencing the burgeoning set of MOOC
resources being generated by the community.  Our solution hybridizes
best practices among ease-of-use descriptions, search practices and
the persistence and identification standards.  Our work aims to
catalyse work towards making linked open data a closer reality for 
the world's learners.

\section*{Acknowledgement}
This research is funded in part by NUS Learning Innovation Fund -- Technology 
grant \#C-252-000-123-001, and also in part by the scholarship from 
China Scholarship Council (CSC), National Natural
Science Foundation of China under Grant Nos.: 61673085 and 61433014,
and UESTC Fundamental Research Funds for the Central Universities
under Grant No.: ZYGX2016J196. We also thank the anonymous reviewers for 
their useful comments.
% \bibliography{mooc}
%

\end{document}